\documentclass[useAMS,usenatbib]{mn2e}
\usepackage{amsmath}
\usepackage{epsfig}
\usepackage{graphicx,psfrag,amssymb}
\newcommand{\figdir}{./}
\usepackage{amssymb}
\newcommand{\mitbf}[1] {\hbox{\mathversion{bold}$#1$}} 
\newcommand{\rmd}{{\rm d}}

\begin{document}

\title{On mode conversion and wave reflection in magnetic Ap stars}
\author[S.~G. Sousa and M.~S. Cunha]{S. G. Sousa$^{1,2}$\thanks{E-mail: sousasag@astro.up.pt} and M. S. Cunha$^{1}$\\
$^{1}$ Centro de Astrof{\'\i}sica da Universidade do Porto, Rua das Estrelas, s/n, 4150-762 Porto, Portugal \\
$^{2}$ Departamento de Matem\'atica Aplicada, Faculdade de Ci\^encias da Universidade do Porto, Portugal}

\date{Accepted ??.  Received ??; in original form 2007 July} \pubyear{2007}

\maketitle

\begin{abstract}
We investigate the effect of a strong large scale magnetic field on the reflection of high frequency acoustic modes in rapidly oscillating Ap stars. To that end we consider a toy model composed of an isothermal atmosphere matched onto a polytropic interior and determine the numerical solution to the set of ideal magneto-hydrodynamic equations in a local plane-parallel approximation with constant gravity. Using the numerical solution in combination with approximate analytical solutions that are valid in the limits where the magnetic and acoustic components are decoupled, we calculate the relative fraction of energy flux that is carried away in each oscillation cycle by running acoustic waves in the atmosphere and running magnetic waves in the interior. For oscillation frequencies above the acoustic cutoff we show that most energy losses associated with the presence of running waves occur in regions where the magnetic field is close to vertical. Moreover, by considering the depth dependence of the energy associated with the magnetic component of the wave in the atmosphere we show that a fraction of the wave energy is kept in the oscillation every cycle. For frequencies above the acoustic cutoff frequency such energy is concentrated in regions where the magnetic field is significantly inclined in relation to the local vertical. Even though our calculations were aimed at studying oscillations with frequencies above the acoustic cutoff frequency, based on our results we discuss what results may be expected for oscillations of lower frequency.

\end{abstract}


\section{Introduction}
A number of rapidly oscillating Ap stars, including the two prototypes $\alpha$~Cir \citep{kurtz94} and  HR1217 \citep{kurtz05}, are known to pulsate with frequencies that exceed the acoustic cutoff frequency expected from appropriate stellar models. This fact has been a matter of long debate over the years \cite[e.g.][]{shibahashi85,audard98,cunha98,gautschy98}. Generally, it has been suggested that the apparent dilemma posed by the observation of oscillations of such high frequency in roAp stars results from the use of a Temperature - Optical depth ($T-\tau$) relation that is inadequate for these peculiar stars. However, except for the case in which the $T-\tau$ relation was modified to simulate the presence of a chromosphere, for which no observational evidence exists, models with modified $T-\tau$ relations considered in the above mentioned works, have not been able to fully resolve the apparent discrepancy. 

Even though the oscillations observed in roAp stars have essentially an acoustic nature, it is nowadays well established that the latter acquire a magnetoacoustic nature in the outer layers of the star, where the magnetic and gas pressure become comparable. A number of studies investigating on the effect of the magnetic field on the properties of the oscillations, such as eigenfrequencies and eigenfunctions, were carried out by different authors \cite[e.g.][]{dziembowski96,bigot00,cunha00,saio04,cunha06}. Mode conversion -- i.e, the process by which the acoustic wave in the interior passes part of its energy to waves of a different nature through the coupling with the magnetic field in the region where the gas and magnetic pressures are comparable -- was considered in all the above mentioned works. In some of them the waves were artificially reflected at the surface of the star. In those cases only the energy passed onto running magnetic waves in the interior was considered. In the studies of \cite{dziembowski96} and \cite{cunha06}, on the other hand, the authors did not make the former assumption, and as a result, part of the energy associated with the acoustic wave in the interior was allowed to be converted, through the coupling with the magnetic field, into running acoustic waves in the high atmosphere, as well as into running magnetic waves in the interior.

While incorporating the process of mode conversion, the works mentioned above aimed mostly at studying the overall effect of the presence of the magnetic field on the eigenfrequencies and eigenfunctions of the oscillations. Thus, in no case the impact that such process might have on the issue of mode reflection at the surface of roAp stars was properly analysed. This issue, which is directly related to the long debated problem of the cutoff frequency in roAp stars discussed at the beginning of this Section, is studied in the present work in a toy model composed of an isothermal atmosphere matched onto an index 3 polytropic interior. We note that the phenomena of mode conversion is known to take place in other classes of pulsating stars. In particular, in recent years a number of studies related to mode conversion were carried out also in the context of solar pulsations in regions of strong magnetic field \cite[e.g.][and references therein]{mcdougall07}.

\section{The magnetic boundary layer}

\subsection{Governing equations and general assumptions}
In this work we are particularly concerned with pulsations in Ap stars. Observations indicate that Ap stars have intense, large scale, magnetic fields, often with predominant dipolar structure \citep[e.g.][]{hubrig05,hubrig04,wade00}. Accordingly, in the present work we will consider that our stellar model is permeated by a dipolar magnetic field. Since this is a force free field, it does not influence the equilibrium structure of the star, which therefore is governed by the system of equations,
\begin{equation}
\frac{\partial\rho_0}{\partial t} = 0 ,
\label{eq2.2.1}
\end{equation}
\begin{equation}
\frac{\partial p_0}{\partial t} = 0 ,
\label{eq2.2.2}
\end{equation}
\begin{equation}
\nabla p_0 = \rho_0 \mitbf{g}_0 , 
\label{eq2.2.3}
\end{equation}
where $\rho_0$ is the density, $p_0$ is the pressure, $\mitbf{g_0}$ is the gravitational field, and the subscript 0 refers to the equilibrium quantities. In the above equations $\partial$ stands for partial derivatives and $t$ is the time.

In the limit of perfect conductivity, adiabatic motions are governed by the set of ideal magneto-hydrodynamic (MHD) equations \cite[see, e.g.,][]{priest82}. Considering small deviations from the equilibrium state and then linearizing the MHD equations by neglecting squares and products of the small quantities (that will be denoted with the subscript 1), and neglecting the Eulerian perturbation to the gravitational potential (Cowling approximation), we arrive to the following system of equations,
\begin{equation}
\frac{\partial \mitbf{v}}{\partial t} = - \frac{1}{\rho_0}\nabla p_1 + \frac{\rho_1}{\rho_0^2}\nabla p_0 + \frac{1}{\mu\rho_0}(\nabla \times \mitbf{B}_1 ) \times \mitbf{B}_0 ,
\label{eq2.2.9}
\end{equation}
\begin{equation}
\frac{\partial p_1}{\partial t} = - \mitbf{v} \cdot \nabla p_0 - \gamma p_0 \nabla \cdot \mitbf{v} ,
\label{eq2.2.10}
\end{equation}
\begin{equation}
\frac{\partial \rho_1}{\partial t} = - \mitbf{v} \cdot \nabla \rho_0 - \rho_0 \nabla \cdot \mitbf{v} ,
\label{eq2.2.11}
\end{equation}
\begin{equation}
\frac{\partial \mitbf{B}_1}{\partial t} = \nabla \times ( \mitbf{v} \times \mitbf{B}_0) ,
\label{eq2.2.12}
\end{equation}
\begin{equation}
\nabla \cdot \mitbf{B}_1 = 0 ,
\label{eq2.2.12b}
\end{equation}
where $\mitbf{v}$ is the pulsation velocity, $\gamma$ is the the first adiabatic exponent, $\mu$ is the magnetic permeability, and ${\mitbf{B}}$ is the magnetic field. The perturbed Lorentz force in equation (\ref{eq2.2.9}) expresses the magnetic field influence on the oscillations.

\begin{figure}
\centering
\includegraphics[scale=0.35]{\figdir/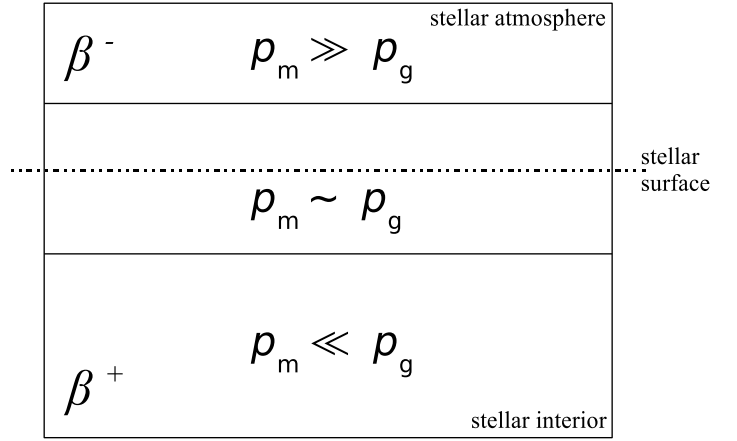}
\caption{Schematic view of the the outer layers of the magnetic star. The two outer regions, where the magnetic pressure is larger than, or of the same order of the gas pressure will be denominated magnetic boundary layer.}
\label{model}
\end{figure}

In our study we will deal only with the outer layers of the star. We shall consider three different regions (cf.\ Fig.~\ref{model}), namely,
\begin{itemize}
\item A region in the atmosphere where the magnetic pressure dominates over the gas pressure ($ \beta^- $ region);
\item A region where the magnetic and the gas pressure are of the same order of magnitude;
\item A region where the magnetic pressure is negligible when compared with the gas pressure ($ \beta^+ $ region).
\end{itemize}
Here $\beta$ is defined as the ratio between the gas pressure, $p_g$, and the magnetic pressure, $p_m = B^2/2\mu$. The region $\beta^-$ corresponds to the layers where $\beta \ll 1$, while the region $\beta^+$ corresponds to the layers where $\beta \gg 1$.

We shall refer to the two outer regions as the {\it magnetic boundary layer}. We note that since in the interior of the star the gradient of the gas pressure is much greater than the Lorentz stresses, the dynamics there is effectively field free. Thus, the magnetic field influences the global oscillations only through its direct contribution to the restoring force in the magnetic boundary layer.

In this study we will consider perturbations to the equilibrium state of the model associated with high order, low degree, axisymmetric pulsations only. These are the pulsations that are typically observed in roAp stars. Moreover, we will not consider pure incompressible perturbations (even though in the $\beta^+$ region the decoupled magnetic component becomes almost incompressible).

Since the magnetic field is assumed to have a dipolar configuration and the pulsations are axisymmetric, the perturbations depend on co-latitude $\theta$ but not on the axial angle $\phi$.

Under these conditions the system of equations (\ref{eq2.2.9})-(\ref{eq2.2.12b}) admits solutions of the form,
\begin{equation}
\bar{\mbox{\boldmath $\xi$}}(r,\theta,t) = {\mbox{\boldmath $\xi$}}(r,\theta)e^{i\omega t} ,
\label{eq2.2.13}
\end{equation}
where $ \bar{\mbox{\boldmath $\xi$}}$ is the displacement vector defined by $\mitbf{v} = \frac{\partial\bar{\mbox{\boldmath $\xi$}}}{\partial t}$, $\omega$ is the oscillation frequency and $r$ is the radial coordinate, in the spherical coordinate system $(r,\theta,\phi)$.

Since the boundary layer is thin and the magnetic field varies only on large scales, we follow \citet[][]{dziembowski96} and \citet[][]{cunha00} and solve our problem locally by adopting, at each latitude, a plane parallel approximation, and assuming a locally uniform field. 

At each given latitude, our Cartesian system is defined such that the local vertical coordinate $z$ increases outwardly and is zero at the surface of the star. The coordinates in the horizontal plane, $x$ and $y$, are chosen such that the $x$ axis is parallel to the horizontal component of the magnetic field. Hence $ \mitbf{B}_0 = (B_x,0,B_z)$.

In accordance with our approximation of a locally uniform magnetic field, we disregard the derivatives of $B_x$ and $B_z$ in equations (\ref{eq2.2.9})-(\ref{eq2.2.12b}). Since the oscillations considered are axisymmetric, $\partial{\mbox{\boldmath $\xi$}}/\partial y = 0$. Moreover, given that the observed modes are of low degree, we expect that ${\mbox{\boldmath $\xi$}}$ will vary much more rapidly in the $z$ direction than in the $x$ direction. Hence, we also approximate $\partial{\mbox{\boldmath $\xi$}}/\partial x \approx 0$. The latter approximation is exact for pulsations that in the absence of the magnetic field would be spherically symmetric (i.e.\ with degree $l=0$).

The presence of a magnetic field, and the consequent Lorentz force that is generated, introduces horizontal motion in the magnetic boundary layer even in the case of modes that in the absence of a magnetic field would be radial. We can decompose the displacement into its vertical component and its horizontal component defined, respectively, by $\xi_z = \mbox{\boldmath $\xi$}\cdot\mitbf{\hat{e}}_z$  and $\xi_x = \mbox{\boldmath $\xi$}\cdot\mitbf{\hat{e}}_x$, where $\mitbf{\hat{e}}_z$ and  $\mitbf{\hat{e}}_x$ are unit vectors in the $z$ and $x$ directions, respectively. The fact that the horizontal displacement is along $\mitbf{\hat{e}}_x$ is a consequence of the axisymmetry of the problem.

Using equations (\ref{eq2.2.10})-(\ref{eq2.2.12}) to eliminate $p_1$, $\rho_1$ and $\mitbf{B}_1$ from equation (\ref{eq2.2.9}), neglecting the derivatives of  $B_x$ and $B_z$, and finally separating into vertical and horizontal components we obtain,
\begin{equation}
-\omega^2\rho_0\xi_x = -\frac{B_x}{\mu} \left[(\mitbf{B}_0\cdot\nabla)(\nabla\cdot\mbox{\boldmath $\xi$}) \right]+\frac{1}{\mu}(\mitbf{B}_0\cdot\nabla)^2 \xi_x ,
\label{eq2.2.14}
\end{equation}
\begin{equation}
\begin{split}
&  -\omega^2\rho_0\xi_z = \frac{\partial W}{\partial z} -\frac{B_z}{\mu} \left[(\mitbf{B}_0\cdot\nabla)(\nabla\cdot\mbox{\boldmath $\xi$}) \right]\\
&  +\frac{1}{\mu}(\mitbf{B}_0\cdot\nabla)^2 \xi_z -\frac{1}{\rho_0}\frac{d p_0}{d z} \nabla \cdot (\rho_0 \mbox{\boldmath $\xi$}) ,
\end{split}
\label{eq2.2.15}
\end{equation}
where,
\begin{equation}
W = \mbox{\boldmath $\xi$} \cdot \nabla p_0 + (\gamma p_0 + \frac{B_0^2}{\mu})\nabla \cdot \mbox{\boldmath $\xi$} - \frac{1}{\mu}(\mitbf{B}_0 \cdot \nabla)(\mitbf{B}_0 \cdot \mbox{\boldmath $\xi$}) .
\label{eq2.2.16}
\end{equation}

\subsection{Dimensionless Variables}
When solving our problem we will make the system of equations (\ref{eq2.2.14})-(\ref{eq2.2.15}) dimensionless by defining the new variables:
$$
\eta = -z/R,\qquad  \sigma = \omega / \omega_0,\qquad p=p_0/\breve{p}_0, 
$$
$$
\rho=\rho_0/\breve{\rho}_0,\qquad \varepsilon_x=\xi_x/R,\qquad \varepsilon_z=\xi_z/R,
$$
$$
b_i=B_{i}/(\mu\breve{\rho}_0\omega_0^2 R^2)^{1/2},\qquad C_2=\breve{p}_0 / (\breve{\rho}_0\omega_0^2 R^2) ,
$$
where $\breve{p}_0$ is the pressure at the stellar surface, $\breve{\rho}_0$ the density at the stellar surface and $\omega_0 = \sqrt{G M / R^3}$, where  $M$ and $R$ are, respectively, the mass and radius of the star, and $G$ is the gravitational constant. Following the same notation, the dimensionless magnetic field vector becomes ${\rm{\mitbf b_0}}=(b_x,0,b_z)$ and the total dimensionless displacement becomes $\bar{\mitbf{\varepsilon}}=\varepsilon_x \mitbf{\hat e}_x+\varepsilon_z \mitbf{\hat e}_z$. Moreover, we define the dimensionless sound speed as $c=(C_2\gamma p/\rho)^{1/2}$.

Combining these new variables with the system of equations (\ref{eq2.2.14})-(\ref{eq2.2.15}) and simplifying, we obtain:
\begin{equation}
-b_x b_z \varepsilon_z''+b_z^2 \varepsilon_x''+\sigma^2 \rho \varepsilon_x = 0 ,
\label{eq2.3.1}
\end{equation}
\begin{equation}
(\gamma C_2 p + b_x^2) \varepsilon_z'' + \gamma C_2 p' \varepsilon_z' + \sigma^2 \rho \varepsilon_z - b_x b_z \varepsilon_x'' = 0 ,
\label{eq2.3.2}
\end{equation}
where the prime $'$ stands for derivatives in order to the new independent variable $\eta$.

\subsection{Magnetic coordinate system}

\begin{figure}
\centering
\includegraphics[scale=0.25]{\figdir/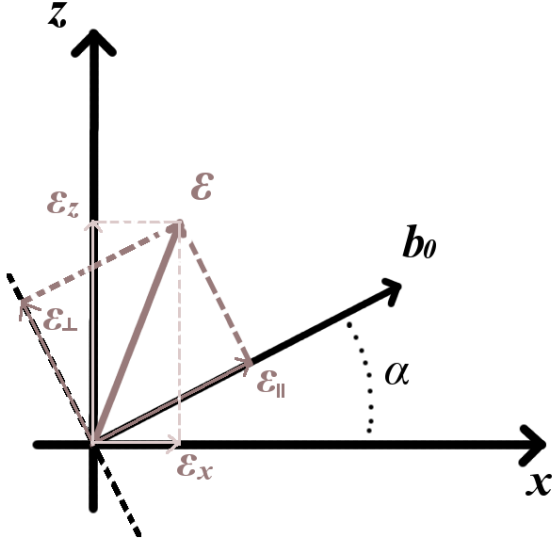}
\caption{Schematic view of the components of the displacement $\mitbf{\varepsilon}$  projected onto the two coordinate systems used in the present work. $\varepsilon_x$ and $\varepsilon_z$ are the dimensionless horizontal and vertical components of the displacement, respectively. $\varepsilon_{\parallel}$ and $\varepsilon_{\perp}$ are the dimensionless components of the displacement along and perpendicular to $\mitbf{b}_0$, respectively. The angle between the local magnetic field direction and the local horizontal coordinate $x$ is designated by $\alpha$.}
\label{coord}
\end{figure}

When the system of equations (\ref{eq2.3.1})-(\ref{eq2.3.2}) is studied in the extreme case of no gas pressure it is found that the displacement is perpendicular to the magnetic field. This is a direct consequence of the Lorentz force being the only restoring force playing a role in that extreme case. With this in mind we will define a second coordinate system with axes directed along and perpendicular to the magnetic field (see Fig.~\ref{coord}). To change from the first to the second coordinate system we use the relation:
\begin{equation}
 \begin{cases}
  \varepsilon_x = \varepsilon_{\parallel} \cos{\alpha} - \varepsilon_{\perp} \sin{\alpha} \\
  \varepsilon_z = \varepsilon_{\perp} \cos{\alpha} + \varepsilon_{\parallel} \sin{\alpha}
 \end{cases} ,
\label{eq2.3.6}
\end{equation}
where $\alpha$ is the inclination angle of the magnetic field in relation to the horizontal direction and $\varepsilon_{\parallel}$ and $ \varepsilon_{\perp}$ are the dimensionless components of the displacement along and perpendicular to the magnetic field direction, respectively. This coordinate system will be useful when dealing with the problem in the $\beta^-$ region. On the other hand it will be useful to use the $(x,z)$ coordinate system in the $\beta^+$ region, since in the region where the gas pressure dominates over the magnetic pressure the vertical component of the displacement is essentially an acoustic wave \citep[e.g.][]{campbell86,cunha00}.

\section{Stellar Model}

The main goal of this work is to understand how high frequency magnetoacoustic pulsations are reflected in the outer layers of Ap stars, and what fraction of the pulsation energy is lost through running waves. Thus, we start by considering a simple model that still contains the main physical ingredients necessary to study this problem, namely a plane parallel envelope under constant gravitational acceleration, composed of an isothermal atmosphere matched at the surface onto a polytropic interior of index 3. We hope to refine our analysis using a more sophisticated stellar model, in future work. 

All results presented are for a model of mass $M=2.0$~M$_\odot$ and radius $R=2.0$~R$_\odot$, where R$_\odot$ and M$_\odot$ are the radius and mass of the sun, respectively. The temperature of the isothermal atmosphere (where the perfect gas law is assumed to hold) is $T=8440$~K and the atmosphere is matched continuously onto the polytropic interior at a pressure $P=7563$~$gcm^{-1}s^{-2}$. These values were chosen by comparison with a main sequence (age~=~468Myrs) stellar model of the same mass and radius computed with the CESAM code \citep{morel97}.

The expressions for the density and pressure profiles are given by \citep[e.g.][]{gough93}
\begin{equation}
\rho={\rm e}^{\frac{\eta}{{\mathcal H}}},
\end{equation}
\begin{equation}
p={\rm e}^{\frac{\eta}{{\mathcal H}}},
\end{equation}
in the isothermal atmosphere, where ${\mathcal H}=H/R$ and $H$ is the scale height given by $H=p_0R^2/(\rho_0 GM)$, and by 
\begin{equation}
\rho=\left(\frac{\eta}{\eta_s}+1\right)^3
\end{equation}
\begin{equation}
p=\left(\frac{\eta}{\eta_s}+1\right)^4,
\end{equation}
in the polytropic interior, where $\eta_s=4{\mathcal H}$.

\begin{figure}
\centering
\includegraphics[scale=0.30]{\figdir/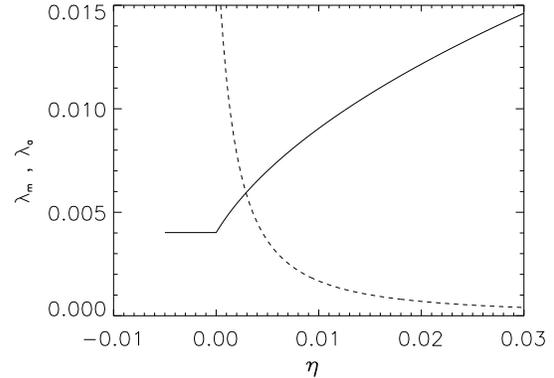}
\caption{Characteristic dimensionless magnetic (dashed line) and acoustic (full line) wavelengths as a function of dimensionless depth in the star, as defined by equations (\ref{eq2.2.30}) and (\ref{eq2.2.29}), respectively. These were derived for the model under study, considering a cyclic frequency $\nu = 3.08$~mHz and a characteristic magnetic field of 3000~G. }
\label{lambdas}
\end{figure}

In Fig.~\ref{lambdas} we show the characteristic (dimensionless) magnetic and acoustic oscillation wavelengths for the model under study defined, respectively, by
\begin{equation}
\lambda_m = \frac{2 \pi |\mitbf{b}_0|}{ \sigma \sqrt{\rho}}
\label{eq2.2.30}
\end{equation}
and
\begin{equation}
\lambda_a = \frac{2 \pi c}{ \sigma },
\label{eq2.2.29}
\end{equation}
assuming a cyclic oscillation frequency $\nu = 3.08$~mHz and a magnetic field strength of 3000 G. In the isothermal atmosphere ($\eta < 0$) the characteristic magnetic wavelength is much greater than the characteristic acoustic wavelength. Deeper in the star there is a region where the two wavelengths are of the same order of magnitude, and deeper still a region where the acoustic wavelength is much greater than the magnetic wavelength. Thus, it is expected that the magnetoacoustic mode described by equations (\ref{eq2.2.14})-(\ref{eq2.2.15}) will decouple onto acoustic and magnetic components, both sufficiently high in the atmosphere and sufficiently deep in the interior of the star, so far as the waves propagate there. The exact regions of decoupling will depend on the magnetic field intensity and on the gas pressure in the outer layers of the star. 

\section{Decoupling regions}

\subsection{$\beta^+$ Region}

As shown by \cite{roberts83}, deep enough in the star the magnetoacoustic wave decouples into its fast acoustic and slow magnetic components, and it is possible to find an approximate solution for the horizontal component of the dimensionless displacement $\varepsilon_x$, by applying appropriate standard asymptotic theory valid for high frequency modes. Within the approximations considered in the present work, the horizontal component of the displacement in the $\beta^+$ region is entirely magnetic and has the form,

\begin{equation}
 \begin{split}
 & \varepsilon_x \approx \rho^{-1/4} D\exp \left\{ -i \int_0^{\eta} \Bigl( \frac{\rho\sigma^2}{b_z^2}\Bigr)^{1/2}{\rm d}\eta\right\} + \\
 &\rho^{-1/4} E \exp\left\{ i \int_0^{\eta} \Bigl( \frac{\rho\sigma^2}{b_z^2}\Bigr)^{1/2}{\rm d}\eta\right\}
 \end{split}
\label{eq3.3.2}
\end{equation}
where $D$ and $E$ are depth independent amplitudes.
Following the same authors we assume that this component of the displacement is an inwardly propagating wave and set  $E = 0$. This is justified by the fact that the latter describes a magnetic wave that is expected to dissipate while propagating inwardly, due to the rapid increase of its wavenumber with depth.

\subsection{\label{beta-}$\beta^-$ Region}
If the oscillation frequency is sufficiently high, the wave will propagate in the atmosphere. In that case we expect the displacement to decouple into a slow acoustic and a fast magnetic component, as a result of the difference in the characteristic magnetic and acoustic wavelengths in this region. Thus, under these conditions the total displacement may be written as the sum of fast (subscript $f$) and slow (subscript $s$) oscillatory components,
\begin{equation}
\mbox{\boldmath $\varepsilon$} = \mbox{\boldmath $\varepsilon$}_f +\mbox{\boldmath $\varepsilon$}_s.
\label{eq2.3.11}
\end{equation}
In the $\beta^-$ region the perturbed Lorentz force dominates the restoring force for perturbations that are perpendicular to the unperturbed magnetic field. Hence, the component of the displacement in the direction perpendicular to the unperturbed magnetic field will be associated essentially with a compressible magnetic wave. However, to first order the Lorentz force has no effect for perturbations along the direction of the unperturbed magnetic field. Thus, the dimensionless displacement along that direction will be associated with a wave that is essentially acoustic. We thus expect to have,
\begin{equation}
\varepsilon_{\parallel s} \gg \varepsilon_{\perp s},
\label{eqEllsggEpds}
\end{equation}
and
\begin{equation}
\varepsilon_{\perp f} \gg \varepsilon_{\parallel f}.
\label{eqEpdfggEllf}
\end{equation}

Rewriting the system of equations (\ref{eq2.3.1})-(\ref{eq2.3.2}) in the coordinate system defined by equations (\ref{eq2.3.6}) we find,
\begin{equation}
\gamma C_2 p \Bigl( \varepsilon_{\perp}'' + \varepsilon_{\parallel}''\frac{\sin {\alpha}}{\cos{\alpha}}\Bigr) + 
\gamma C_2 p' \Bigl( \varepsilon_{\perp}' + \varepsilon_{\parallel}'\frac{\sin {\alpha}}{\cos{\alpha}}\Bigr) + \frac{\sigma^2\rho}{\sin {\alpha} \cos{\alpha}} \varepsilon_{\parallel} = 0 ,
\label{eq2.3.9}
\end{equation}
\begin{equation}
- \frac{b_x b_z}{\cos{\alpha}} \varepsilon_{\perp}'' + \sigma^2\rho \Bigl( \varepsilon_{\parallel} \cos{\alpha} - \varepsilon_{\perp} \sin{\alpha} \Bigr) = 0 .
\label{eq2.3.10}
\end{equation}
Since the coefficients multiplying the displacement in equations (\ref{eq2.3.9})-(\ref{eq2.3.10}) are not wave like functions, these equations must be satisfied also by the fast and slow components of the displacement separately. Neglecting $\varepsilon_{\perp s}$ when compared with $\varepsilon_{\parallel s}$ and $\varepsilon_{\parallel f}$ when compared with $\varepsilon_{\perp f}$, we find that the equations governing the slow and fast components of the displacement are, within the referred approximations, 
\begin{equation}
\varepsilon_{\parallel s}'' + \frac{p'}{p} \varepsilon_{\parallel s}' + \frac{\sigma^2 \rho}{\gamma C_2 p \sin^2\alpha} \varepsilon_{\parallel s} = 0
\label{eq2.3.20}
\end{equation}
and
\begin{equation}
\varepsilon_{\perp f}'' + \frac{p'}{(p+\hat{B}^2)} \varepsilon_{\perp f}' + \frac{\sigma^2 \rho}{\gamma C_2 \cos^2\alpha (p+\hat{B}^2)} \varepsilon_{\perp f} = 0,
\label{eq2.3.23}
\end{equation}
respectively, where
\[ \hat{B}^2 = \frac{|\mitbf{b}_0|^2}{\gamma C_2 \cos^2 \alpha}. \]
This decoupling of the magnetoacoustic wave into acoustic and magnetic components in the $\beta^-$ region will be verified numerically in Section~\ref{sec:sol}.
Equations (\ref{eq2.3.20}) and (\ref{eq2.3.23}) may be written in the form of a standard wave equation. To that end we define the new variable
\begin{equation}
\Xi_{\parallel s} = p^{1/2} \varepsilon_{\parallel s}
\label{eq2.3.25}
\end{equation}
and introduce it in equation (\ref{eq2.3.20}) to find,
\begin{equation}
\Xi_{\parallel s}'' + k_{\parallel}^2 \Xi_{\parallel s} = 0 ,
\label{eq2.3.26}
\end{equation}
where
\begin{equation}
k_{\parallel}^2 = \Biggl[ \biggl(\frac{ p'}{2 p}\biggr)^2 - \frac{1}{2} \frac{p''}{p} + \frac{\sigma^2 \rho}{\gamma C_2 p \sin^2 \alpha}\Biggr] .
\label{eq2.3.33}
\end{equation}
For equation (\ref{eq2.3.23}) we can use a similar variable, defined by
\begin{equation}
\Xi_{\perp f} = (p + \hat{B}^2)^{1/2} \varepsilon_{\perp f} ,
\label{eq2.3.27}
\end{equation}
to find,
\begin{equation}
\Xi_{\perp f}'' + k_{\perp}^2 \Xi_{\perp f} = 0 ,
\label{eq2.3.28}
\end{equation}
where
\begin{equation}
k_{\perp}^2 = \Biggl[ \biggl(\frac{p'}{2 (p + \hat{B}^2) }\biggr)^2 - \frac{1}{2} \frac{p''}{(p+\hat{B}^2)} + \frac{\sigma^2 \rho}{\gamma C_2 (p+ \hat{B}^2) \cos^2 \alpha}\Biggr].
\label{eqkpd}
\end{equation}
Equations (\ref{eq2.3.20}) and (\ref{eq2.3.23}) were derived under the assumption that two decoupled wave components with significantly different characteristic wavelengths can be identified in the $\beta^-$ region. That condition is satisfied when $k_{\parallel}^2$ and $k_{\perp}^2$ are both positives. 
$k_\parallel^2$ is positive for oscillations with frequencies larger than $\sigma_c$ where,
\begin{equation}
\sigma_c = \frac{c}{2\mathcal{H}} \sin \alpha .
\label{eq2.3.37}
\end{equation}
Except for the dependence in the angle $\alpha$, this expression is similar to the cutoff frequency for acoustic waves in an isothermal atmosphere with no magnetic field (hereafter named {\it  acoustic cutoff frequency}). In fact, when the magnetic field is vertical the local vertical motion associated with the radial pulsations (or a low degree mode) will not be affected and we recover the acoustic cutoff frequency. As the magnetic field becomes more inclined in relation to the local vertical, the necessary frequency for propagation of acoustic waves in the atmosphere decreases.

\begin{figure}
\centering
\includegraphics[scale=0.30]{\figdir/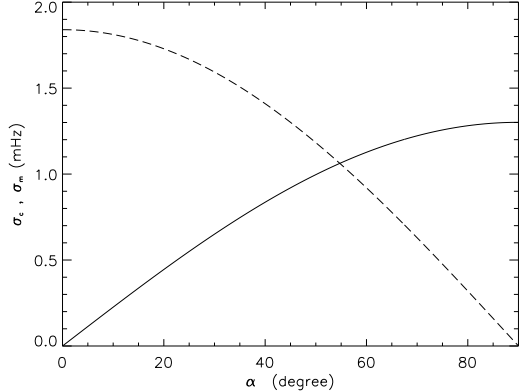}
\caption{The acoustic (full line) and magnetic (dashed line) frequencies defined by equations (\ref{eq2.3.37}) and (\ref{eq2.3.40}), respectively, for our model, and their dependence on the inclination of the magnetic field in relation to the local horizontal direction, $\alpha$.}
\label{critical}
\end{figure}

Concerning the magnetic component we find that $k_{\perp}^2$ is positive for oscillations with frequencies larger than $\sigma_m$, where
\begin{equation}
\sigma_{m} = \frac{c}{\mathcal{H}} \cos \alpha \left[\frac{1}{2} - \frac{1}{4\left(1+\frac{\hat{B}^2}{p}\right)}\right]^{1/2}.
\label{eq2.3.39}
\end{equation}
The frequency $\sigma_m$ increases with height in the atmosphere, taking a maximum value (as $\eta \rightarrow -\infty$) of
\begin{equation}
\hat\sigma_{m} = \frac{\sqrt{2}c}{2\mathcal{H}} \cos \alpha .
\label{eq2.3.40}
\end{equation}
In  Fig.~\ref{critical} we show the frequencies $\sigma_c$ and $\hat\sigma_{m}$for the model used in the present work.

We emphasize that equations (\ref{eq2.3.20}) and (\ref{eq2.3.23}) are approximately valid only in the outer layers, 
 where the magnetic and acoustic components of the wave are decoupled. Hence, they may not be used to directly infer under which conditions the magnetoacoustic wave is reflected (or partially reflected) in the layers below. For that reason we avoided using the term `critical frequencies' when referring to $\sigma_c$ and $\sigma_m$. In section~\ref{sec:energy} we shall look at the problem of wave reflection and pulsation energy losses more closely, using the numerical solutions obtained for the magnetic boundary layer.

The components of the displacement in the $\beta^-$ region can be obtained from the solutions to equations (\ref{eq2.3.20}) and (\ref{eq2.3.23}).   
 The general solution to equation (\ref{eq2.3.26}) is given by
\begin{equation}
\Xi_{\parallel s} = \tilde{A} e^{i k_{\parallel} \eta} + \tilde{B} e^{-i k_{\parallel} \eta}
\label{eq3.3.1}
\end{equation}
where $\tilde{A}$ and $\tilde{B}$ are depth independent amplitudes.
Since no energy can be sent in from outside the star, this must represent a wave propagating outwardly through the isothermal atmosphere. Thus $\tilde{B}$ must be zero. Using equation~(\ref{eq2.3.25}) we thus find the solution to equation~(\ref{eq2.3.20}) to be,
 \begin{equation}
\varepsilon_{\parallel s} = \frac{\tilde A}{p^{1/2}}e^{i k_{\parallel} \eta}.
\label{waveacoustic}
\end{equation}

To find an approximate solution to equation~(\ref{eq2.3.23}) we first note that when terms of order $\beta^2$ are neglected, $k_{\perp}^2$ becomes
\begin{equation}
k_{\perp}^2\approx\frac{\rho}{|{\mitbf b_0}|^2}\left(\sigma^2-\hat\sigma_{m}^2\right).
\end{equation}
Using $\rho$ as the independent variable, equation~(\ref{eq2.3.28}) thus becomes,
\begin{equation}
\rho^2\frac{\rmd^2 \Xi_{\perp f}}{\rmd\rho^2}+\rho\frac{\rmd\Xi_{\perp f}}{\rmd\rho}+\rho\frac{\mathcal{H}^2}{{|{\mitbf b_0}|^2}}\left(\sigma^2-\hat\sigma_{m}^2\right)\Xi_{\perp f} = 0.
\label{magwave}
\end{equation}

The solution to equation~(\ref{magwave}) can be written in terms of Bessel functions $J_0$ and $Y_0$ \cite[e.g.][pg 362]{abramowitz72}, namely,
\begin{equation}
\Xi_{\perp f} = \tilde C J_0\left(2\sqrt\mathcal X\rho\right) + \tilde D Y_0\left(2\sqrt\mathcal X\rho\right),
\end{equation}
where $\tilde C$ and $\tilde D$ are depth independent amplitudes and $\mathcal{X}={\mathcal{H}^2}{{|{\mitbf b_0}|^{-2}}}\left(\sigma^2-\hat\sigma_{m}^2\right)$. Since $Y_0$ diverges as $\rho \rightarrow 0$, $\tilde D$ must be zero. Thus, using equation~(\ref{eq2.3.27}), we find that within the approximations described, the solution to equation~(\ref{eq2.3.23}) has the form,
\begin{equation}
\varepsilon_{\perp f} = \frac{\tilde C}{\left(p+\hat B^2\right)^{1/2}} J_0\left(2\sqrt\mathcal X\rho\right).
\label{magfinal}
\end{equation} 

\section{Solutions in the magnetic boundary layer}
\label{sec:sol}

Throughout most of the magnetic boundary layer, the displacement is associated with a magnetoacoustic wave which cannot be described, not even  approximately, by decoupled magnetic and acoustic components. Thus, to determine the form of the displacement in this case it is necessary to solve the system of equations (\ref{eq2.3.1})-(\ref{eq2.3.2}) numerically applying, simultaneously, appropriate boundary conditions.  In the absence of a magnetic field, the acoustic cutoff frequency is equal to the polar value of $\sigma_c$. Since our goal is to investigate on the reflection of oscillations of which frequency is above this acoustic cutoff frequency, in what follows we will consider the case of a cyclic frequency well above $c/(2\pi 2\mathcal{H})$.

\subsection{Boundary Conditions}
Since the system of equations to be solved is linear, in order to solve it we need to apply three boundary conditions. Two of these conditions will be applied in the $\beta^-$ region while the third will be applied in the $\beta^+$ region. 

At the outermost point of the atmosphere we match the perturbed magnetic field into a vacuum field. This condition implies that $\varepsilon_{\perp}'= 0$ \citep[e.g.][]{cunha06}.
To derive the two additional boundary conditions we use the fact that two approximate analytical solutions can be obtained, one in each of the decoupling regions.
In the $\beta^-$ region the parallel displacement, $\varepsilon_\parallel$, is essentially associated with the acoustic wave. Hence, in this region the parallel component of the numerical solution must be well described by the analytical expression given by equation~(\ref{waveacoustic}).
Moreover, in the $\beta^+$ region the horizontal component of solution is essentially associated with the magnetic wave. Hence in this region the horizontal component of the numerical solution must be well described by  
the asymptotic solution given by equation (\ref{eq3.3.2}) with $E = 0$.

To find a solution that obeys both the system of equations (\ref{eq2.3.1})-(\ref{eq2.3.2}) and the boundary conditions given above we start by finding three linearly independent particular solutions that satisfy the system of equations and the first of these boundary conditions. To generate these three particular solutions we had to provide starting conditions that assured they were linearly independent. The first, second and third solutions were generated by assuming at the outermost point of integration, respectively, ($\varepsilon_{\perp} = 1$, $\varepsilon_{\parallel}= 0$, $\varepsilon_{\parallel}'= 0$),  ($\varepsilon_{\perp} = 0$, $\varepsilon_{\parallel}= 1$, $\varepsilon_{\parallel}'= 0$) and ($\varepsilon_{\perp} = 0$, $\varepsilon_{\parallel}= 0$, $\varepsilon_{\parallel}'= 1$).  To apply the two additional boundary conditions we write the total solution as a linear combination of these particular solutions,

\begin{equation}
{\mitbf{\varepsilon}} = {\mitbf{\varepsilon}_1} + c_1 {\mitbf{\varepsilon}_2} + c_2 {\mitbf{\varepsilon}_3}
\label{eq3.3.3}
\end{equation}

\begin{figure*}
\centering
\includegraphics[scale=0.42]{\figdir/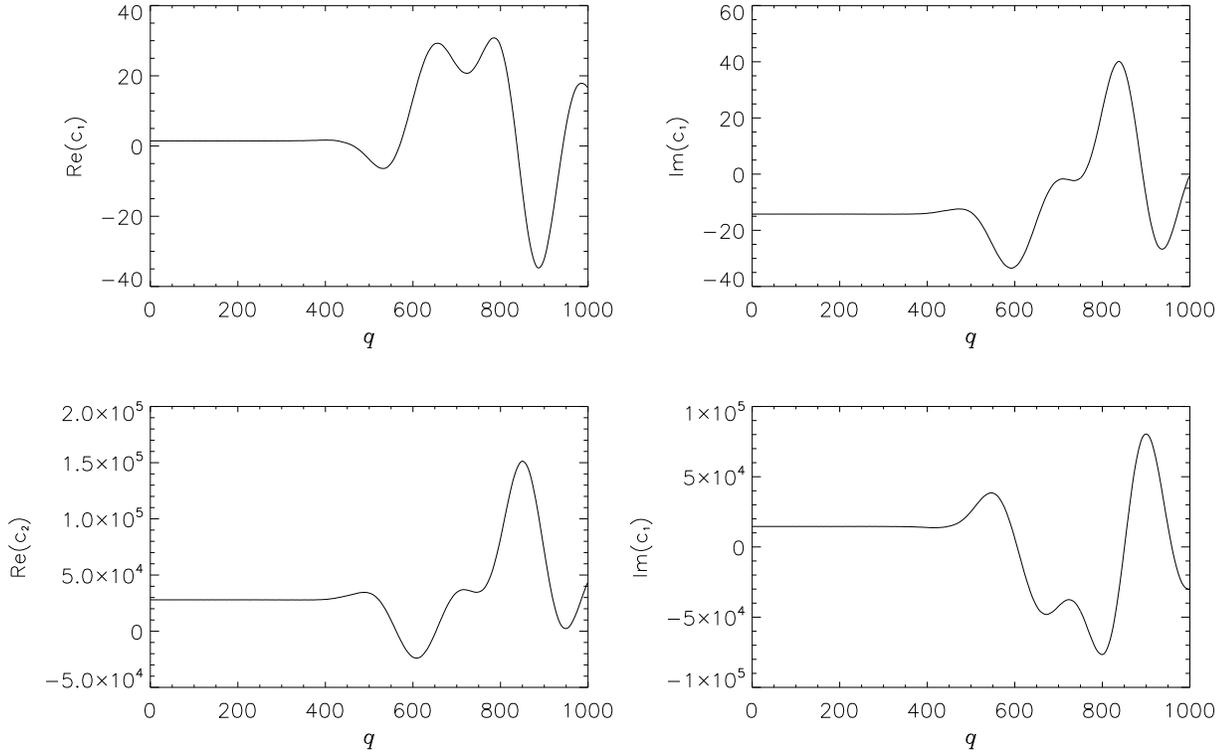}
\caption{The figure shows the behaviour of the complex coefficients of the linear combination expressed by equation~(\ref{eq3.3.3}). A high value of the index {\it q} means that the  matchings are carried out in a region where $\beta\sim 1$, while low values of that index mean that the matchings are carried out in the regions $\beta^+$ and $\beta^-$.  $c_1$ and $c_2$ tend to constant values as the index {\it q} tends to zero, confirming the decoupling of the magnetoacoustic components both in the outer atmosphere and in the interior. Results are for a dipolar  magnetic field with polar strength $B_p = 3000$~G at a latitude such that its inclination with respect to the local horizontal axis is $\alpha = 45^{\circ}$, and a cyclic frequency $\nu=3.08$~mHz.}
\label{const50}
\end{figure*}

The complex coefficients $c_1$ and $c_2$ are determined by applying the two remaining boundary conditions, i.e. by matching simultaneously the relevant components of the numerical solution to the analytical solution (\ref{eq3.3.2}) with $E=0$ in the deeper layers and to the analytical solution (\ref{waveacoustic}) in the atmosphere. When performing these matchings we introduced an integer index {\it q} which reflects the positions at which the matchings are carried out  in a grid of depths. The value of this index decreases from a thousand to zero as the depths at which the matchings are carried out are moved from the region where $\beta\sim 1$ to the regions where the magnetic and acoustic wave components are decoupled (see Appendix A for details). 
As mentioned in the previous Section, we can check the decoupling of the magnetoacoustic wave in the $\beta^-$ and $\beta^+$ regions numerically. If such decoupling occurs, the coefficients $c_1$ and $c_2$ must tend to constant values as the index {\it q} tends to zero.

Fig.~\ref{const50} shows the behaviour of $c_1$ and $c_2$ when the matchings are carried out at different depths. The results shown are for an oscillation with cyclic frequency $\nu=3.08$~mHz and a dipolar magnetic field of polar strength $B_p=3000$~G at a latitude such that its inclination with respect to the local horizontal axis is $\alpha = 45^{\circ}$. When the matchings are carried out in the region where $\beta \sim 1$, which in Fig.~\ref{const50} corresponds approximately to {\it q} ranging between 600 and 1000, the coefficients $c_1$ and $c_2$ vary significantly with the index {\it q}. This is because the analytical solutions used in the matchings are not good representations of the true solutions in the regions considered.
However, as we move the matchings simultaneously to the regions $\beta^-$ and $\beta^+$ (close to $q=0$) $c_1$ and $c_2$ tend to constant values, confirming the expected decoupling of the magnetoacoustic wave in both regions.

\begin{figure*}
\centering
\includegraphics[scale=0.44]{\figdir/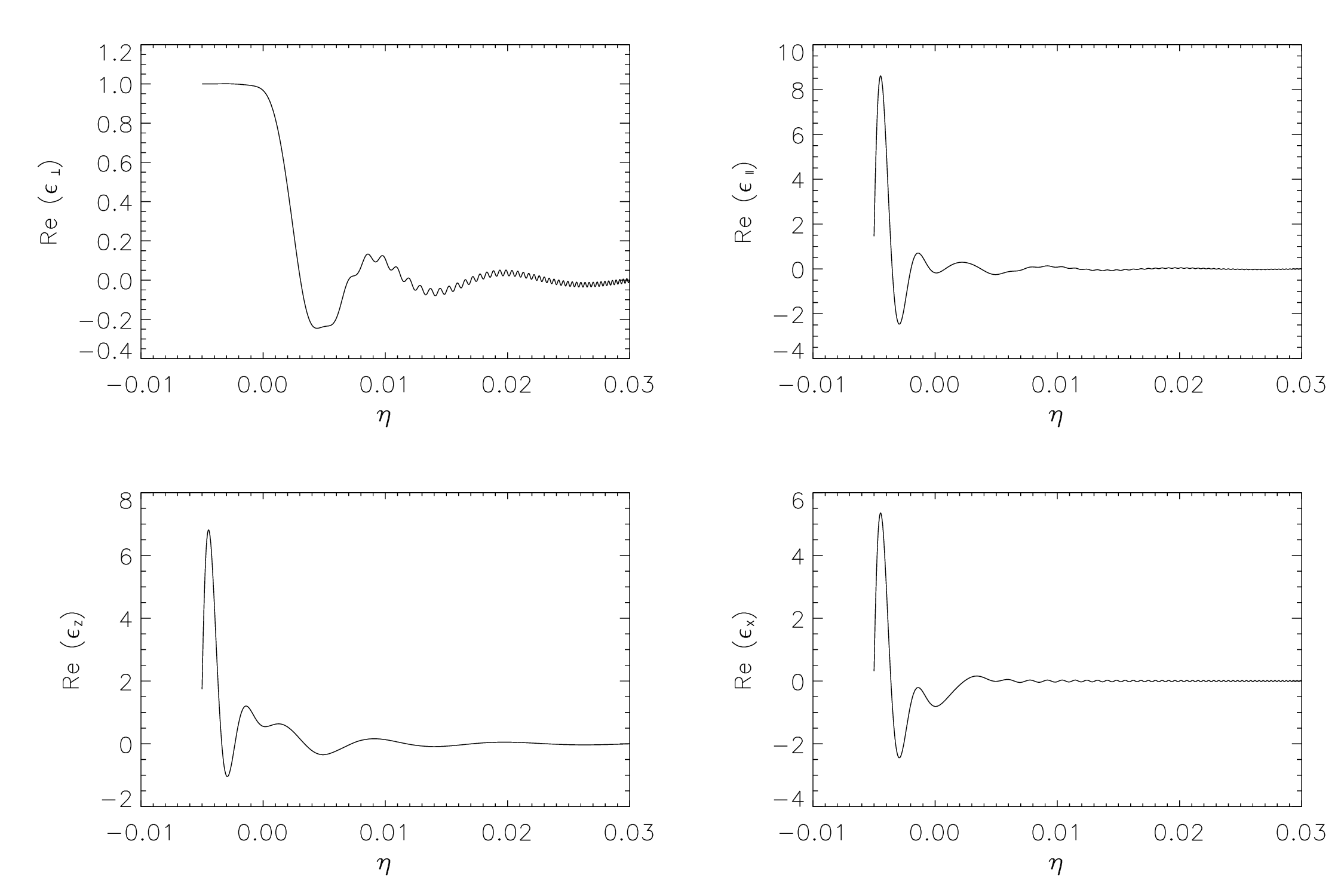}
\caption{The real parts of the dimensionless displacement perpendicular to the magnetic field (top left), along the magnetic field (top right), in the vertical direction (bottom left), in the horizontal direction (bottom right). Results are for a dipolar  magnetic field with polar strength $B_p = 3000$~G at a latitude such that its inclination with respect to the local horizontal axis is $\alpha = 45^{\circ}$, and a cyclic frequency $\nu=3.08$~mHz.}
\label{solucoes50}
\end{figure*}

\subsection{Numerical Solutions}

Having calculated the constants for the linear combination we can determine the total solution for the displacement. For illustration we show in Fig.~\ref{solucoes50} the real part of dimensionless displacement for a dipolar magnetic field of polar strength $B_p = 3000$~G, at a latitude such that $\alpha = 45^{\circ}$, and a cyclic oscillation frequency $\nu= 3.08$~mHz.
The figure shows the components of the dimensionless displacement perpendicular and parallel to the direction of the magnetic field (upper panels) and the components of the dimensionless displacement in the vertical and horizontal directions (lower panels).

\begin{figure}
\centering
\includegraphics[scale=0.4]{\figdir/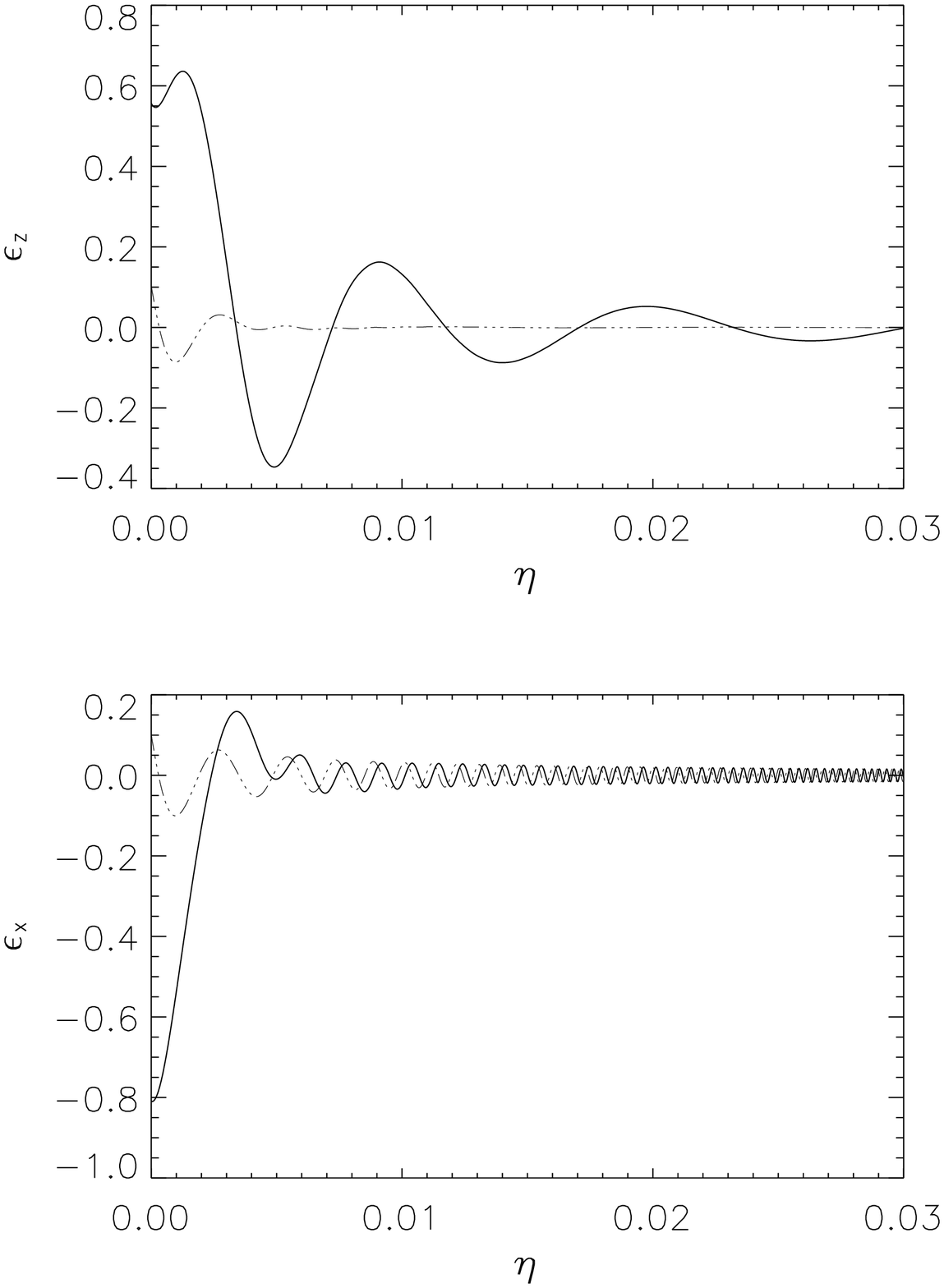}
\caption{The dimensionless displacement in the interior. The top panel shows the vertical component while the lower panel shows the horizontal component. The full and dashed-dotted lines show, respectively, the real and imaginary parts of the displacement. Results are for the same magnetic field and the same frequency as in Fig.~\ref{solucoes50}.}
\label{sol50ZUbmais}
\end{figure}

\begin{figure}
\centering
\includegraphics[scale=0.4]{\figdir/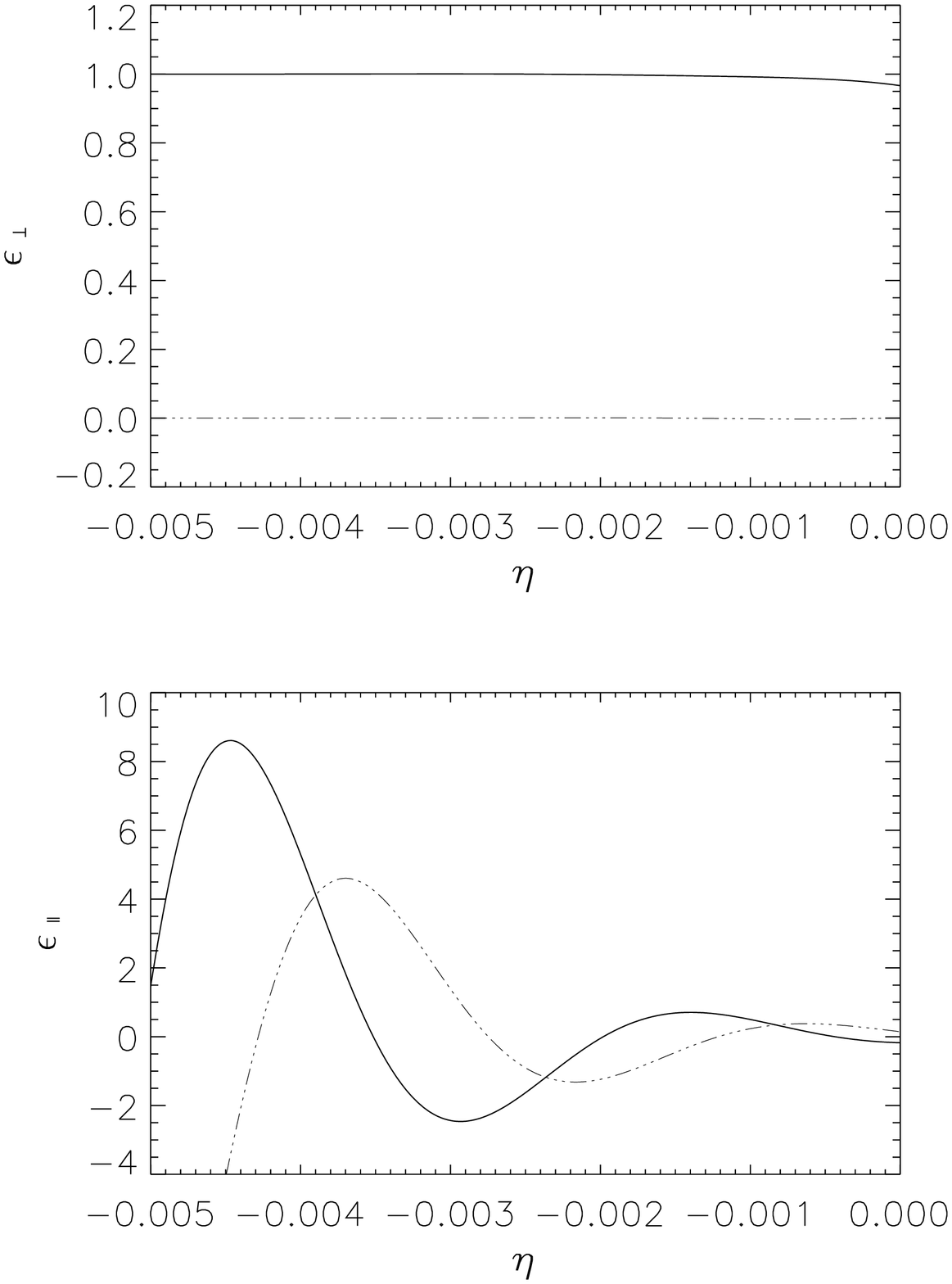}
\caption{The dimensionless displacement in the atmosphere. The top panel shows the component perpendicular to the magnetic field while the lower panel shows the component along the magnetic field. The full and dashed-dotted lines show, respectively, the real and imaginary parts of the displacement. Results are for the same magnetic field and the same frequency as in Fig.~\ref{solucoes50}.}
\label{sol500EpdEllbmenos}
\end{figure}

The decoupling of the magnetoacoustic wave into acoustic and magnetic components both in the $\beta^+$ and $\beta^-$ regions can be seen in the plots. To illustrate that, zooms of Fig.~\ref{solucoes50} in the interior and in the atmosphere are shown in Fig.~\ref{sol50ZUbmais} and Fig.~\ref{sol500EpdEllbmenos}, respectively. Fig.~\ref{sol50ZUbmais} shows the vertical (upper panel) and horizontal (lower panel) solutions in the interior. The decoupled acoustic and magnetic waves are readily seen. The horizontal component is a rapidly varying magnetic oscillation, while the vertical component is an acoustic oscillation with a significantly larger wavelength. Moreover, inspection of the real and imaginary parts of the horizontal displacement confirm the running nature of the magnetic wave in the interior, characterized by real and imaginary parts of similar amplitude and out of phase by $\pi/2$. The nearly standing nature of the acoustic wave in the interior is also clear from the imaginary part of the vertical displacement, which amplitude tends to zero with increasing depth (alternatively, a standing wave could have real and imaginary parts with significant amplitude, but in phase).
Fig.~\ref{sol500EpdEllbmenos} shows the component of the dimensionless displacement perpendicular to the magnetic field direction (upper panel) and the component of the same vector along the magnetic field direction (lower panel), in the atmosphere. As before, the acoustic and magnetic components are easily identified. The component along the direction perpendicular to the magnetic field tends to a constant, as expected for the magnetic wave from the analysis carried out in Section~\ref{beta-}, while the component of the dimensionless displacement along the direction of the magnetic field is an acoustic oscillation with a wavelength comparable to the thickness of the atmospheric layer shown. Once again, inspection of the real and imaginary parts of the displacement confirm the running nature of the acoustic wave in the atmosphere, and the standing nature of the magnetic wave in the same region. We note that although the standing nature of the magnetic wave in the atmosphere is connected to the condition of matching of the perturbed magnetic field onto a vacuum field, the form of this component of the displacement is rather insensitive to the place at which that boundary condition is applied, so far as the latter is applied sufficiently high in the atmosphere. Also, as shall be discussed in Section~\ref{sec:energy}, this nature of the magnetic wave is not associated to a full wave reflection at a particular place in the atmosphere, but rather to a progressive decrease of its energy content with height.

Fig. \ref{solucoes50}, \ref{sol50ZUbmais}, and \ref{sol500EpdEllbmenos} illustrate also the fact that in the atmosphere the decoupling takes place in the parallel and perpendicular directions, while in the interior it takes place in the horizontal and vertical directions, instead.

\section{Energy Fluxes}
\label{sec:energy}
\subsection{Flux carried by running waves}
The acoustic waves that propagate outwardly in the atmospheres of roAp stars  and the magnetic waves that propagate inwardly in their interiors, both take energy away from the global oscillations. This is true both for oscillations with frequencies below and above the acoustic cutoff frequency. For oscillations with frequencies below the acoustic cutoff frequency, energy losses through running waves in the atmosphere are restricted to regions where the inclination of the magnetic field is such that $\sigma > \sigma_c$. On the other hand, oscillations with frequencies above the acoustic cutoff are expected to lose energy through running waves in the atmosphere regardless of the inclination of the magnetic field.  For this reason the observation in some roAp stars of oscillations with frequencies above the acoustic cutoff frequency has been a matter of long debate. At first sight, one would be led to think that perturbations of such high frequency would simply propagate away in the atmosphere and be dissipated before they could grow to observable amplitudes. However, the simple fact that a global oscillation of such high frequency is observed indicates that only part of the mode energy is removed by the running acoustic and magnetic waves. Moreover, it also indicates that the energy input to such a mode in each cycle, through the opacity mechanism acting in the region where hydrogen is ionized \citep{balmforth01,cunha02,saio05}, is sufficient to compensate all sources of mode energy losses, including the energy removed by the running waves. 

With the question of mode reflection in mind, in this Section we estimate the fraction of energy flux that is carried away by running waves and its dependence on the inclination of the magnetic field. As before, the calculations are carried out in a local plane-parallel approach, assuming a dipolar magnetic field and using the numerical solutions for the displacement obtained in Section \ref{sec:sol} to compute the relevant energy fluxes at each co-latitude.

\begin{figure}
\centering
\includegraphics[scale=0.38]{\figdir/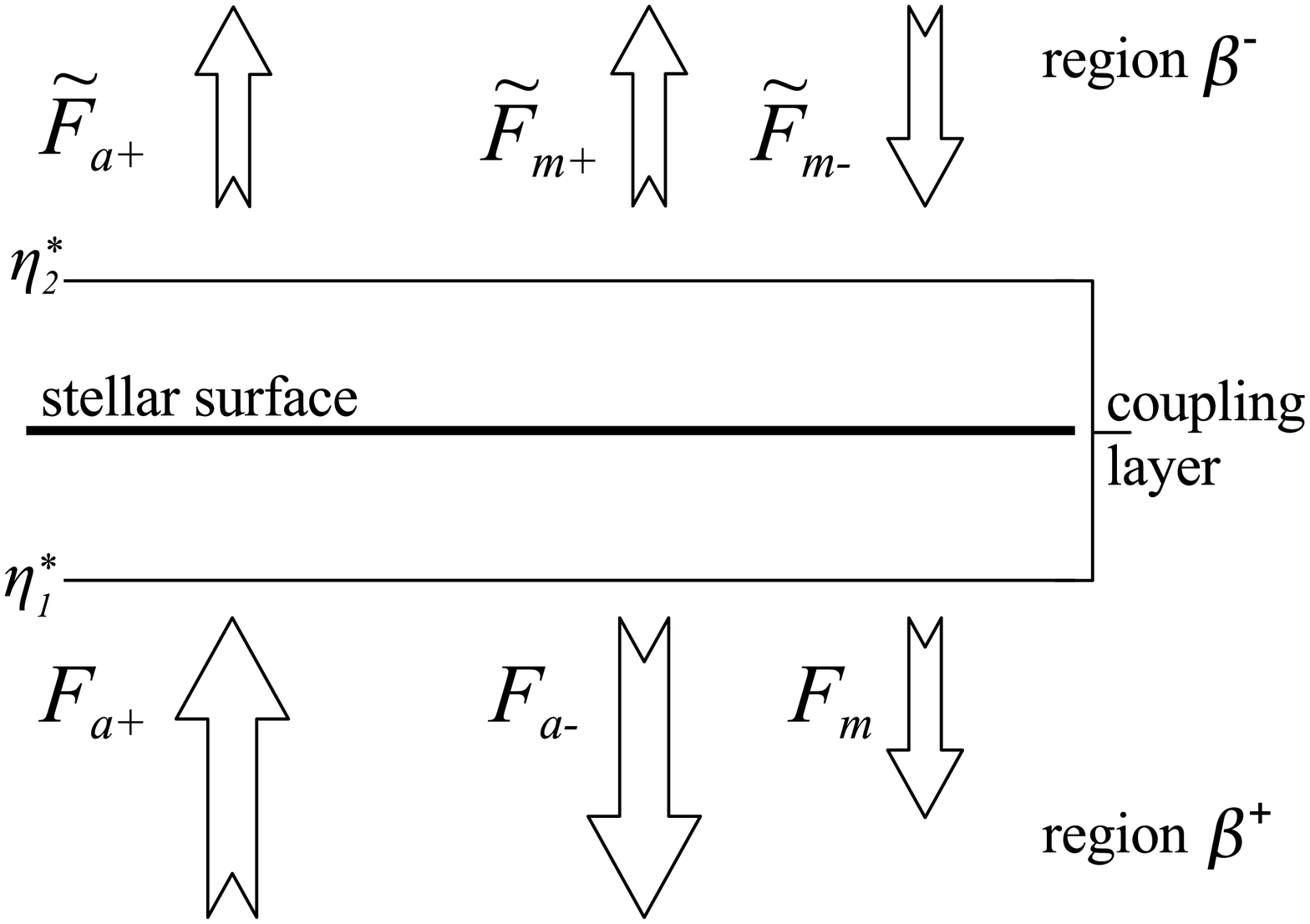}
\caption{Sketch of the energy fluxes carried by the waves below and above the coupling region.}
\label{reflections}
\end{figure}

Let us consider a wave that propagates outwardly from the interior of the star carrying a given energy flux (see Fig.~\ref{reflections}). At some point the wave is reflected, or partially reflected, and propagates inwardly again, back to the starting point. As a result of the magnetoacoustic coupling near the surface layers, this wave transfers part of its energy to a inwardly propagating magnetic wave which eventually dissipates in the $\beta^+$ region. Moreover, if the oscillation frequency is greater than the frequency defined by equation~(\ref{eq2.3.37}), part of the wave energy is transferred to an outwardly propagating acoustic wave.

Since the layer where the acoustic and the magnetic components are coupled is thin, and the imaginary part of the frequency is expected to be very small \cite[e.g.][]{cunha99}, the amplitude of the oscillation changes only very slightly during the time that takes the wave to propagate through the coupling layer. Therefore, using energy conservation expressed in terms of the dimensionless variables we can write, approximately,

\begin{equation}
F_{a+}(\eta^*_1,\tau_0) - F_{a-}(\eta^*_1,\tau_0) \approx  \tilde{F}_{a+}(\eta_2^*,\tau_0)+ F_{m}(\eta^*_1,\tau_0)
\label{eq4.1.2}
\end{equation}
where $\tau_0=t_0\omega_0$ and $t_0$ is the time at which the outwardly propagating wave crosses the height $\eta = \eta^*_1$. Moreover, $F_{a+}$, $F_{a-}$, $\tilde{F}_{a+}$ and $F_{m}$ are the dimensionless energy fluxes averaged over one oscillation period carried in the decoupled regions by the outwardly propagating acoustic component below the coupling layer, the inwardly propagating acoustic component in the same region, the outwardly propagating acoustic wave above the coupling layer, and the inwardly propagating magnetic wave in the interior, respectively. Here the dimensionless energy fluxes are defined as the corresponding energy fluxes divided by the quantity $R^3\omega_0^3\breve\rho_0$.

The average dimensionless energy flux $F_{a+}$ can be calculated using the asymptotic solution for the dimensionless vertical displacement in the $\beta^+$ region given by \citep[e.g.][]{gough93},

\begin{equation}
\varepsilon_z \approx J \frac{k^{1/2}}{\rho^{1/2}}\cos \Bigl(-\int_{\eta}^{\eta^*_1} k {\rm d}\eta + \delta \Bigr)
\label{eq4.1.8}
\end{equation}
where $J$ is the depth independent amplitude, $\delta$ a complex phase and $k=R\kappa$ with $\kappa$ the vertical wavenumber of the acoustic oscillation in the interior. Using this asymptotic form of the solution, \cite{cunha00} found for the outwardly propagating wave in the interior,
\begin{equation}
F_{a+}(\eta^*_1,\tau_0) \approx \frac{|J|^2\sigma_r^2 c k_r}{8} e^{-2\delta_i} e^{-2\sigma_i \tau_0}
\label{eq4.1.9}
\end{equation}
where the subscripts $r$ and $i$ stand for the real and imaginary parts of the correspondent quantities.

To determine the fraction of flux that is carried away by the running waves, one needs to find expressions for the energy flux associated with the magnetic wave in the interior and the acoustic wave in the atmosphere. An expression for the former was derived by \cite{cunha00} using the asymptotic solution for the magnetic wave given by equation (\ref{eq3.3.2}). From the latter, the authors found,
\begin{equation}
F_m(\eta^*_1,\tau_0) \approx  |D|^2\frac{\sigma_r^2}{2} |{b_z}| e^{-2\sigma_i \tau_0}
\label{eq4.1.11}
\end{equation}
where $D$ is the amplitude of the magnetic wave defined in equation (\ref{eq3.3.2}). 

The flux $\tilde{F}_{a+}$, on the other hand, was not considered by the same authors, who adopted a fully reflective boundary condition at the surface. To derive an expression for the latter, we write the average dimensionless energy flux over one period $P$ carried by the outwardly propagating acoustic wave above the coupling layer in the form,
\begin{equation}
\tilde{F}_{a+}(\eta^*_2,\tau_0) \approx \frac{2}{P} \int_{\tau_0}^{\tau_0+P} \frac{1}{2} \rho |v^+|^2 c {\rm d}\tau
\label{eq4.1.5}
\end{equation}
where $v^+$ is the dimensionless velocity of the oscillation, and the dimensionless group velocity of the wave was approximated by $c$.
Using equation~(\ref{waveacoustic}), we write the solution for the outwardly propagating acoustic wave in the $\beta^-$ region in the form,
\begin{equation}
\bar{\varepsilon}_{\parallel s}^+ = \varepsilon_{\parallel s}^+e^{i\sigma \tau} = \frac{\tilde{A}}{p^{1/2}} e^{i k_{\parallel} \eta} e^{i\sigma \tau},
\label{eq4.1.3}
\end{equation}
and, thus,
\begin{equation}
|v^+|=\Re \left(\frac{\partial \bar{\varepsilon}_{\parallel s}^+}{\partial \tau}\right) \approx \frac{|\tilde{A}|\sigma_r }{p^{1/2}} e^{-\sigma_i \tau} \sin (k_{\parallel} \eta + \sigma_r  \tau ), 
\label{eq4.1.6}
\end{equation}
where we have neglected the small imaginary part of the frequency in the amplitude, when differentiating the displacement.
The corresponding dimensionless energy flux averaged over one period then becomes,
\begin{equation}
\tilde{F}_{a+}(\eta_2^*,\tau_0) \approx \frac{|\tilde{A}|^2 \sigma_r^2 \rho c}{2p} e^{-2\sigma_i \tau_0}
\label{eq4.1.7}
\end{equation}

\begin{figure}
\centering
\includegraphics[scale=0.45]{\figdir/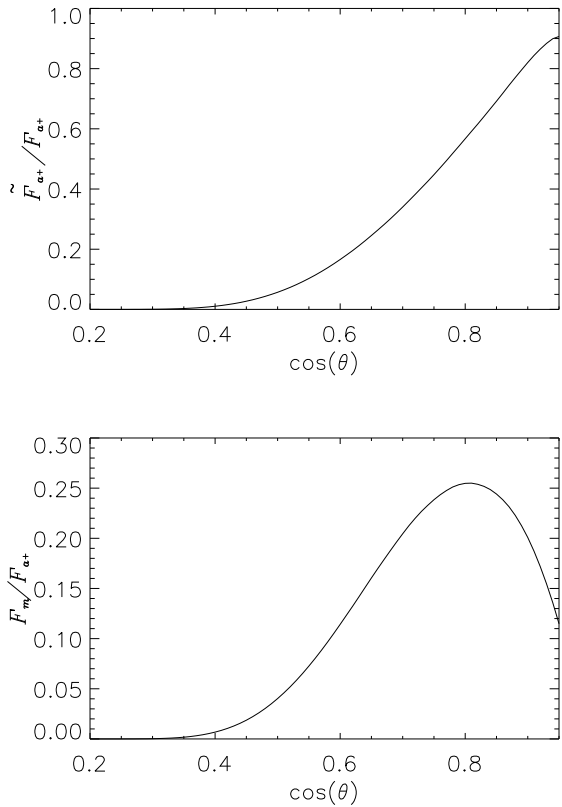}
\caption{Relative energy flux carried by the running acoustic wave in the atmosphere (upper panel) and by the running magnetic wave in the interior (lower panel), for different co-latitudes. The results shown are for a dipolar magnetic field with polar strength $B_p=3000$~G and a cyclic oscillation frequency $\nu=3.08$~mHz.}
\label{fluxsfafmfa}
\end{figure}

Fig.~\ref{fluxsfafmfa} shows the fraction of energy flux carried by the running acoustic wave in the atmosphere and by the running magnetic wave in the interior. The amplitudes $|J|$, $|D|$ and $|\tilde{A}|$ were obtained from the numerical solutions. The results shown are for a dipolar magnetic field with polar strength $B_p=3000$~G and for a cyclic oscillation frequency $\nu=3.08$~mHz. Note that in this case there is a direct relation between the inclination of the magnetic field and the co-latitude $\theta$, the field being in the local vertical direction when cos~$\theta =1$ and in the local horizontal direction when cos~$\theta =0$. As expected, the fraction of the energy flux carried by the acoustic wave is large near the pole, where the magnetic field is nearly vertical. In the limit of a vertical magnetic field a low degree oscillation in this local analysis does not couple with the magnetic field. Hence, if the frequency is above the acoustic cutoff, as is the case for the example shown, all energy is lost. However, as the inclination of the magnetic field in relation to the local vertical increases, the fraction of energy carried away by the running acoustic wave in the atmosphere diminishes, becoming negligible when the magnetic field becomes strongly inclined.
The fact that no coupling takes place between the magnetic field and the oscillation when the magnetic field is vertical should be reflected also on the inwardly propagating magnetic wave in the interior. In fact, due to the absence of coupling no energy flux is expected to be carried away by this wave when cos~$\theta =1$.  Unfortunately, numerical problems show up when the total fraction of energy flux carried by the running waves approaches 1. Hence we had to stop our calculations at cos~$\theta = 0.95$. Still, inspection of the lower panel of Fig.~\ref{fluxsfafmfa} shows that indeed the fraction of energy flux carried by the running magnetic wave in the interior decreases sharply as the field tends to vertical. Moreover, the energy flux carried away by this magnetic wave increases as the magnetic field becomes slightly more inclined, and then decreases again. A similar result was found by  \cite{cunha00}, who have shown that the exact position in co-latitude of the maximum of energy flux carried by the running magnetic wave in the interior depends on the frequency of the oscillation.

One interesting aspect to note in Fig.~\ref{fluxsfafmfa}, is that the sum of the energy fluxes carried by the running waves is rather large (close to 1) for $\cos\theta > 0.8$. In practice this indicates that for frequencies above the acoustic cutoff, a significant fraction of the energy input in each cycle through the opacity mechanism is lost predominantly in the regions where the magnetic field is close to vertical.

\subsection{ Reflection of the magnetic wave in the atmosphere}

As mentioned above, for frequencies above the acoustic cutoff a significant fraction of energy flux is lost near the magnetic poles through running acoustic waves in the atmosphere and running magnetic waves in the interior. Nevertheless, the fact that global waves with frequencies above the acoustic cutoff frequency are observed in some roAp stars indicates that at least part of the mode energy is conserved in the oscillation over each cycle. In this context it is important to notice that the amplitude of the magnetic component of the wave in the atmosphere at a given co-latitude is approximately constant (e.g.~Fig.~\ref{sol500EpdEllbmenos}). Consequently, the total energy of this wave component must decrease with height, which means that the wave is progressively reflected.  To check this idea we divide the magnetic wave in the atmosphere in outwardly and inwardly propagating components, similar to what was done for the acoustic wave in the interior. 

\begin{figure}
\centering
\includegraphics[scale=0.45]{\figdir/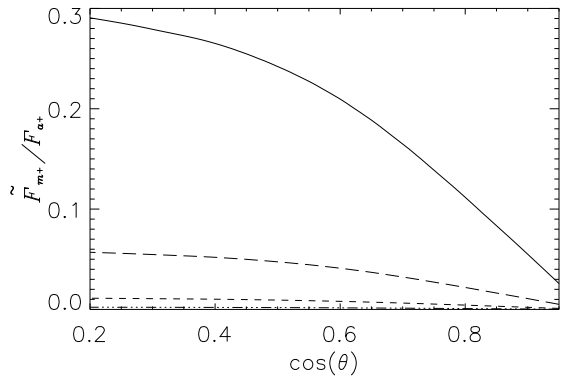}
\caption{Relative energy flux carried by the outwardly propagating component of the magnetic wave in the atmosphere for different co-latitudes and at different heights in the atmosphere, namely, $\eta=-0.001$ (full line),  $\eta=-0.003$ (long dashed line), $\eta=-0.005$ (short dashed line),   $\eta=-0.007$ (dashed-dotted line). The results are for the same magnetic field and oscillation frequency as in Fig.~\ref{fluxsfafmfa}.}
\label{fmtilantes}
\end{figure}

The expression for the outwardly propagating component can be derived from equation~(\ref{magfinal}). In the limit where $\rho \ll 1$ this expression becomes,
\begin{equation}
\bar\varepsilon_{\perp f}^+ = \varepsilon_{\perp f}^+e^{i \sigma \tau}  \approx \frac{\tilde{C}}{2|\hat{B}|}\left(e^{i\sqrt{\frac{2\left(\sigma^2-\hat\sigma_m^2\right)\mathcal{H}^2}{|b_0|^2}} \sqrt{\rho} + \sigma \tau}  \right).
\label{A5}
\end{equation}
Since this wave is essentially a magnetic wave, the energy flux it carries while propagating outwardly averaged over one oscillation period is given by the time average of the Poynting flux. Thus, in terms of the dimensionless variables, we have,
\begin{eqnarray}
\tilde{F}_{m +} (\eta_2^*, \tau_0) & \approx & 
%\hspace{-2.5cm} 
\frac{1}{P} \int_{\tau_0}^{\tau_0+P} \left[\left(\Im\left(\sigma\bar\varepsilon_{\perp f}^+ \hat{\mitbf{\rm e}}_\perp\right) \times \mitbf{b}_0\right)\right.  \nonumber \\
&  &\left.\times \left(\nabla \times \left(\Re\left(\bar\varepsilon_{\perp f}^+ \hat{\mitbf{\rm e}}_\perp\right) \times \mitbf{b}_0\right)\right)\right]_{\bf \hat{e}_z}{\rm d}\tau.  
\label{A6}
\end{eqnarray}
Combining equations (\ref{A5}) and (\ref{A6}) and neglecting, as before, the small imaginary component of the frequency when appropriate, we find,
\begin{equation}
\tilde{F}_{m +} (\eta_2^*, \tau_0) \approx \frac{|\tilde{C|}^2 \sigma_r\left(\sigma_r^2-\hat\sigma_m^2\right)^{1/2} |\mitbf{b}_0|}{8 \hat{|B|}^2} \frac{\sqrt{2}}{2} \rho^{1/2} e^{-2\sigma_i \tau_0}, 
\label{A7}
\end{equation}
where again the amplitude $|\tilde{C}|$ is to be determined by comparison with the numerical solutions.
From equation~(\ref{A7}) it is immediately clear that as the outwardly running component of the magnetic wave travels in the atmosphere, the energy flux it carries decreases due to the factor $\rho^{1/2}$. Hence, the total energy contained in the magnetic wave in the atmosphere will also decrease with height. Fig.~\ref{fmtilantes} shows the fraction of energy flux carried by the outwardly component of the magnetic wave in the atmosphere for different heights in the atmosphere. Besides the rapid decline in the fraction of energy flux carried by this component with increasing height, which results from the $\rho^{1/2}$ factor discussed above, it is clear that most of the energy passed onto the magnetic component in the atmosphere is concentrated in co-latitudes such that $\cos\theta < 0.8$.

\section{Discussion}
It is nowadays well accepted that pulsations in roAp stars are essentially acoustic in nature. Nevertheless, in the outer layers, where the magnetic and gas pressures are comparable, the direct effect of the magnetic field on the oscillations, through the perturbed Lorentz force, becomes important and the oscillations acquire a  magnetoacoustic nature. Away from this region, both above, where the magnetic pressure dominates, and below, where the gas pressure dominates, the magnetoacoustic wave decouples into acoustic and magnetic components. Nevertheless, the presence of the coupling layers results in energy being transferred from the acoustic wave in the interior into waves that are essentially magnetic or acoustic in the decoupled regions, a process commonly known as mode conversion.

In the present work we have used a toy model composed of an isothermal atmosphere matched onto a polytropic interior to investigate on the effect that mode conversion may have on the reflection of oscillations near the surface of roAp stars and also to determine the conditions under which energy losses due to running waves are expected to be most important. 
For frequencies above the acoustic cutoff, we have found that the maximum energy losses due to  running acoustic waves in the atmosphere and running magnetic waves in the interior takes places in regions where the magnetic field is not significantly inclined, which for the dipolar magnetic field considered here, happens around the magnetic poles.

The primary motivation of the current study was the understanding of the presence of oscillations with frequencies above the acoustic cutoff in some roAp stars. Hence, the study did not consider oscillations with frequencies below that limit. Nevertheless, from the present results and the results found by previous studies such as those of \cite{cunha00,saio05,cunha06}, which considered full reflection of the waves in the atmosphere, one can easily extrapolate the results that would be obtained when considering frequencies below the acoustic cutoff. In fact, for oscillations  with frequency below the acoustic cutoff, the acoustic waves in the atmosphere will propagate only if the magnetic field is sufficiently inclined in relation to the local vertical. Thus, the results for oscillation with such frequencies are expected to resemble those found here, except for the fact that the fraction of energy carried away by the running acoustic wave in the atmosphere will decrease sharply to zero when the magnetic field inclination in relation to the local vertical becomes smaller than the value needed to ensure the propagation of the acoustic waves in the atmosphere. Thus, in such case, the maximal energy loss due to the presence of running acoustic waves should take place, not at the magnetic pole, but in an annulus around the latter, where the inner radius of that annulus is defined by the minimum inclination of the magnetic field needed for the propagation of the acoustic waves in the atmosphere.

Despite the energy losses associated with the running waves discussed above, our results show that part of the energy is kept in the oscillation every cycle. In fact, the energy associated with the magnetic wave in the atmosphere decreases significantly with height, which means that the wave is progressively reflected. That energy, which is kept in the oscillation, is concentrated in regions where the magnetic field is significantly inclined in relation to the local vertical, i.e., around the magnetic equator. Naturally, for oscillations with frequencies below the acoustic cutoff, additional energy would be kept in the oscillation around the magnetic poles, where the acoustic component of the atmospheric wave would be evanescent.

The present study considered adiabatic pulsations only, and it is beyond our scope to analyse the impact of the energy losses discussed here on the excitation of the oscillations in roAp stars. \cite{balmforth01} carried out linear non-adiabatic calculations of pulsations in roAp stars imposing a transmissive boundary condition  at the top when the oscillation frequencies were above the acoustic cutoff. However, in their calculations the authors did not consider the direct effect of the magnetic field on pulsations, which means they ignored both the energy associated with the magnetic wave in the atmosphere, which is kept in the oscillation, and the energy losses associated with the magnetic wave in the interior. On a more recent study, \cite{saio05} considered linear non-adiabatic pulsations in models of roAp stars taking into account the direct effect of the magnetic field on the oscillations. However, the author imposed a fully reflective outer bounder condition, which means he ignored the energy losses associated with running acoustic waves in the atmosphere. Clearly what is needed is a study of non-adiabatic pulsations in models of roAp stars which takes into account all the aspects discussed here, including the direct effect of the magnetic field on the oscillations and an outer boundary condition that allows the acoustic waves to propagate away whenever their frequencies are above the frequency $\sigma_c$ defined in equation~(\ref{eq2.3.37}). By providing a better insight of the process of mode conversion and energy exchange in a simple model of the outer layers of roAp stars, we hope the present work can serve as a motivation for the progress of the non-adiabatic calculations described above. 

\section*{Acknowledgments} 
MSC is very grateful to Z.E.~Musielak for fruitful discussions. 
SS and MSC are supported by FCT and FEDER (POCI2010) through the project POCI/CTE-AST/57610/2004FCT-Portugal, and the grant SFRH/BD/17952/2004. The EC's FP6, FCT and FEDER (POCI2010) supported this project also through the HELAS international collaboration.

\appendix{\label{appA}}
\section{Matching procedure}

Since our calculations are linear, the amplitude of the solution to equations~(\ref{eq2.3.1})-(\ref{eq2.3.2}) cannot be determined. Thus, we carry out the matchings necessary to apply the outer and inner boundary conditions by considering the relations,
\begin{equation}
\frac{\varepsilon_{\parallel}'}{\varepsilon_{\parallel}} = i k_{\parallel} - \frac{p'}{2p} \quad \mbox{with } k_{\parallel}^2 > 0,
\label{eq3.3.4}
\end{equation}
and
\begin{equation}
\frac{\varepsilon_x'}{\varepsilon_x} = -i \left(\frac{\rho \sigma^2}{b_z^2}\right)^{1/2},
\label{eq3.3.6}
\end{equation}
 derived, respectively, from the solutions expressed by equation~(\ref{waveacoustic}) and equation~(\ref{eq3.3.2}), the first of which is valid in the $\beta^-$ region and the second of which is valid in the $\beta^+$ region.

\begin{figure}
\includegraphics[scale=0.27]{\figdir/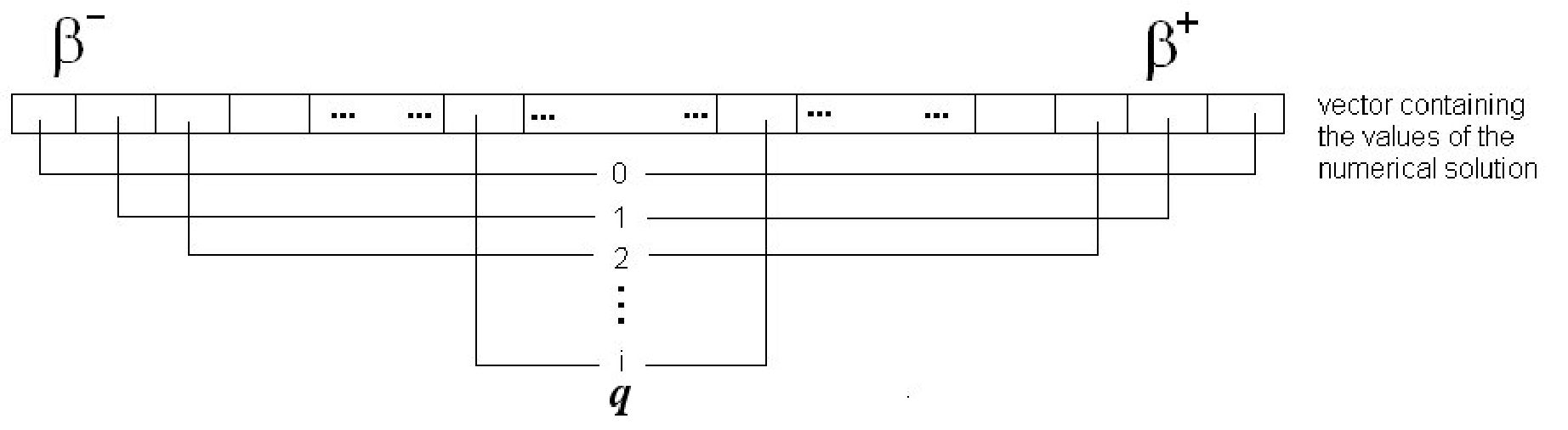}
\caption{Grid of depths used in the matchings. A given value of the index $q$ is assigned to a given pair of depths, such that when the pair of depths is in the region $\beta \sim 1$, $q$ is large, and when the pair of depths includes a point in the $\beta^+$ region and a point in the $\beta^-$ region, $q$ is small.} 
\label{match}
\end{figure}

To compute the left hand side of the expressions given above, we consider the three independent solutions computed numerically, and combine them according to the relations,
\begin{equation}
\frac{\varepsilon_{\parallel}'}{\varepsilon_{\parallel}} = \frac{{\varepsilon_{\parallel 1}}' + c_1 {\varepsilon_{\parallel 2}}' + c_2 {\varepsilon_{\parallel 3}}'}{{\varepsilon_{\parallel 1}} + c_1 {\varepsilon_{\parallel 2}} + c_2 {\varepsilon_{\parallel 3}}}
\label{eqa}
\end{equation}
\begin{equation}
\frac{\varepsilon_x'}{\varepsilon_x} = \frac{\varepsilon_{x1}'+c_1\varepsilon_{x2}'+c_2\varepsilon_{x3}'}{\varepsilon_{x1}+c_1\varepsilon_{x2}+c_2\varepsilon_{x3}}
\label{eqb}
\end{equation}
From these matchings we determine the values of $c_1$ and $c_2$. In practice the matchings expressed by equations~(\ref{eq3.3.4})-(\ref{eq3.3.6}) are carried out for all depths, rather than only in the $\beta^-$ and $\beta^+$ regions. The depths are included in a grid with index $q$, as illustrated in Fig.~\ref{match}.  When $\beta \sim 1$ the index $q$ is close to one thousand. As the index $q$ decreases, the matching expressed by equation~(\ref{eq3.3.4}) is moved towards the $\beta^-$ region and, simultaneously, the matching expressed by equation~(\ref{eq3.3.6}) is moved to the $\beta^+$ region. Since the analytical solutions used in the matchings are valid only when $q$ is small, only for relatively small values of $q$ one may expect to find $c_1$ and $c_2$ constant. These constants are then used to find the total solution for the displacement across the whole magnetic boundary layer, through the relation expressed in equations~(\ref{eq3.3.3}).

\end{document}